\documentclass[prb,amsmath,amssymb,reprint,]{revtex4-1}

\usepackage{graphicx}
\usepackage{dcolumn}
\usepackage{bm}
\usepackage{braket}
\usepackage[euler]{textgreek}

\usepackage[utf8]{inputenc}
\usepackage[T1]{fontenc}
\usepackage{mathptmx}
\usepackage{xcolor}

\begin{document}

\title[]{Probing local pressure environment in anvil cells with nitrogen vacancy (NV\textsuperscript{-}) centers in diamond}

\author{Kin On Ho}
\thanks{These authors contributed equally to this work.}
\author{Man Yin Leung}
\thanks{These authors contributed equally to this work.}
\author{Yaxin Jiang}
\thanks{These authors contributed equally to this work.}
\author{Kin Pong Ao}
\author{Wei Zhang}
\author{King Yau Yip}
\author{Yiu Yung Pang}
\author{King Cho Wong}
\affiliation{
Department of Physics, The Chinese University of Hong Kong,\\ Shatin, New Territories, Hong Kong, China
}

\author{Swee K. Goh}
 \email{skgoh@cuhk.edu.hk}
\author{Sen Yang}
 \email{syang@cuhk.edu.hk}
\affiliation{
Department of Physics, The Chinese University of Hong Kong,\\ Shatin, New Territories, Hong Kong, China
}

\affiliation{
Shenzhen Research Institute, The Chinese University of Hong Kong,\\ Shatin, New Territories, Hong Kong, China
}

\date{\today}

\begin{abstract}
Important discoveries have frequently been made through the studies of matter under high pressure. The conditions of the pressure environment are important for the interpretation of the experimental results. Due to various restrictions inside the pressure cell, detailed information relevant to the pressure environment, such as the pressure distribution, can be hard to obtain experimentally. Here we present the study of pressure distributions inside the pressure medium under different experimental conditions with NV\textsuperscript{-} centers in diamond particles as the sensor. These studies not only show a good spatial resolution, wide temperature and pressure working ranges, compatibility of the existing pressure cell design with the new method, but also demonstrate the usefulness to measure with these sensors as the pressure distribution is sensitive to various factors. The method and the results will benefit many disciplines such as material research and phase transitions in fluid dynamics.

\end{abstract}

\maketitle

\section{\label{sec:level1}Introduction}
Pressure plays an essential role in modern material research. It provides an opportunity to change the sample properties without introducing additional chemical inhomogeneity to the sample. For example, in the heavy fermion intermetallic CePd\textsubscript{2}Si\textsubscript{2} and the iron-based system BaFe\textsubscript{2}As\textsubscript{2}, superconductivity can be induced by pressure. \cite{Mathur98, Paglion10} Furthermore, for many systems, pressure is the only way to reach certain quantum states. In H\textsubscript{3}S, superconductivity with a critical temperature $T_{c}$ of  203~K has been reported when it is pressurized to 155~GPa.\cite{Drozdov15} More recently, a superconducting state with a remarkably high $T_{\rm c}$ of 250 -- 260~K has been reported in LaH$_{10-\delta}$ at around 200~GPa,\cite{Russell,Drozdov2019} as well as $T_{\rm c}$ of 243~K in yttrium hydrides.\cite{Eremets19} In these hydrides, the stabilization of the high $T_{\rm c}$ phase with chemical doping has thus far not been realized. Apart from stabilizing new quantum phases, pressure can also be used to tune various physical phenomena, such as surface plasmon resonance, \cite{Rodriguez19} and glass transitions. \cite{Roland07} Since pressure is one of the key parameters in understanding the physical properties of the materials, a quantitative analysis of the pressure distribution is necessary.

\begin{figure}[t]
    \centering
        \includegraphics[width=0.45\textwidth]{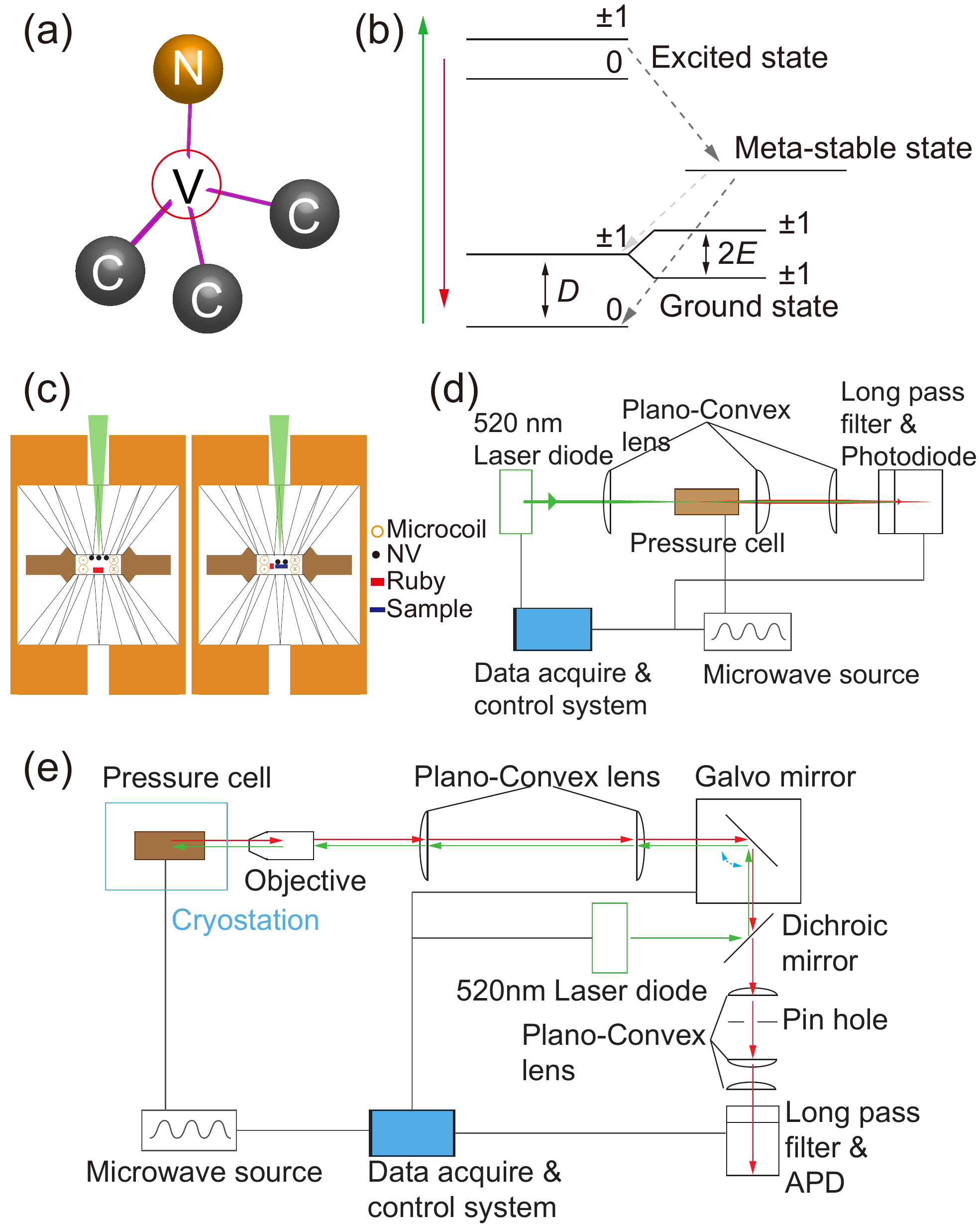} 
    \caption{(a) The structure of NV\textsuperscript{-} center. In a single crystal diamond, the nitrogen substitute could replace one of the four carbon atoms around the vacancy, giving four possible orientations relative to one of the four axes of the carbon bonding. (b) The simplified energy structure of NV\textsuperscript{-} center. \textit{D} ~= 2.87 GHz at ambient pressure and room temperature and \textit{E} is from negligible small to several tens of MHz depending on the diamond sample. (c) The pressure cell consists of a pair of opposing moissanite anvils. The central hole of the gasket is 400 \textmu m with sample, micro coil, NV\textsuperscript{-} centers, and ruby inside. In order to probe the local pressure, NV\textsuperscript{-} centers are drop-casted either on the top of one anvil surface or on the surface of a dummy sample. (d) Fluorescence microscopy setup. A plano-convex lens is used to focus the 520 nm laser to the pressurized region. Given that the diamond particles are very dense, so in principle, the average effects from numerous NV\textsuperscript{-} centers are measured. (e) Confocal microscopy setup. The spatial scanning is achieved by using a galvo mirror. Using the confocal microscope the local pressure experienced by individual NV\textsuperscript{-} centers can be measured.}
\label{fig1}
\end{figure}

Anvil-type pressure cells are commonly employed to reach high pressures. In an anvil cell, the high pressure is reached via mechanically pressing two anvils towards a tightly confined space where the sample and the pressure medium are located, as shown in Fig.~\ref{fig1} (c). There are various possible selections for pressure medium such as glycerin, methanol-ethanol, and helium. When the pressure is below a critical value, the medium stays in the hydrostatic condition, the pressure inside is spatially uniform; once the pressure is above the critical value, the solidification process starts and the pressure distribution inside becomes unavoidably inhomogeneous. Since most of the material characterization methods in high pressure instrumentations probe the response from bulk samples, the interpretation of the results have to be compromised without knowing the detailed distribution of the inhomogeneous pressure. Sometimes, without knowing the exact pressure conditions, contradictory experimental results have been reported. For example, the superconductivity of CaFe$_2$As$_2$ was induced under high pressure in organic pressure media, \cite{Canfield08,Thompson08} while it was not observed at a similar pressure range when helium was used as the medium. \cite{Luke09} This discrepancy may be attributed to the uniaxial stress components that arise from the solidification of the pressure media. \cite{Niklowitz10,Haga09,Tou09,Luke09,Canfield15}

The demanding experimental conditions to reach a high pressure impose a great limitation on the design of the instrument, as well as the available methods for probing pressure distribution \textit{in-situ}. The traditional way to determine pressure in pressure cells is by the optical spectrum of ruby (Cr doped Al\textsubscript{2}O\textsubscript{3}). \cite{Block72,Block73} The optical spectrum of Cr\textsuperscript{3+} near 694 nm contains two resonances: \textit{R}$_1$ and \textit{R}$_2$. Both resonances are sensitive to the pressure (0.364 \AA/kbar) and temperature (0.068 \AA/K). \cite{Eremets96} Usually the pressure is calibrated at room temperature. Although in cryogenic conditions, calibration can be done by adding a pressure insensitive temperature sensor, it is somewhat harder to use ruby spectroscopy due to the drop of fluorescence. \cite{Kawamura00,Mizuki12,Jin00,Schiferl92,Aoyama08,Nicol92} Spatial resolution is also limited due to the relatively large size of ruby and hence the limited number of ruby particles that can be placed inside the pressure chamber. In this work, we present a systematic study of using point defects in diamond particles as the quantum sensor to probe the pressure distribution in pressure medium with high spatial and pressure resolutions. We will also demonstrate the means of tracking the solidification process of pressure medium in wide pressure and temperature ranges.

Nitrogen vacancy (NV) center is a point defect in diamond with a substitutional nitrogen atom and an adjacent lattice carbon vacancy. With an additional electron NV becomes NV\textsuperscript{-} which is a spin 1 system. 
High fidelity optical initialization and spin state readout make NV\textsuperscript{-} center a good quantum sensor. Recently extensive research works show that it can serve as a highly sensitive and spatially-resolved sensor for various physical parameters such as temperature, \cite{Lukin13, Wrachtrup13, Guo11, Budker10, Hollenberg14} electric field, \cite{Wrachtrup11, Yao18} and magnetic field. \cite{Yang18, Prozorov18, Roch18, Yao19} Its performance retains robustness in a wide temperature range \cite{Guo11, Budker10, Hollenberg14} and even under pressure up to 60 GPa. \cite{Prawer14} Furthermore, NV\textsuperscript{-} centers can be used as a stress tensor sensor. \cite{Tetienne19} This motivates us to investigate NV\textsuperscript{-} center under pressure and develop it to be a new type of pressure sensor.

The NV\textsuperscript{-} center has a strong crystal field around it which is sensitive to the distortion of the crystal lattice. The resulting longitudinal zero-field splitting (ZFS) \textit{D} splits $\Ket{0}$ and $\Ket{\pm 1}$ states, and it is due to the first-order electron spin-spin interaction. The origin of the pressure dependence is the change of distances of the electron spins. This is attributed to the global compression and additional local structural relaxation, but is not related to the change of the spin density distribution on the neighbour shells. \cite{Prawer14, Gali14} The transverse ZFS \textit{E}, which is related to the electric field or internal strain (see Appendix for more discussions), can further split the degenerate $\Ket{\pm 1}$ state to two superposition states of $\Ket{\pm 1}$ (eigenstates of $S_{x,y}$) with an energy difference 2\textit{E}.

As shown in Fig.~\ref{fig1} (b), NV\textsuperscript{-} center has its particular electronic structure and spin state dependent inter-system crossing which leads to the optical spin initialization, readout and spin state dependent fluorescence rate. Therefore the electron spin resonance (ESR) can be measured via the optically detected magnetic resonance (ODMR) method. \cite{Wrachtrup1997} The longitudinal ZFS \textit{D} can be extracted from the center frequency of the resonances in ODMR spectra. From that, we can derive the pressure (details in section III).

\section{The apparatus and experimental protocol}

The protocol of using NV\textsuperscript{-} centers to measure pressure has been proposed and demonstrated before. \cite{Prawer14} In that work, a piece of diamond sample was placed inside the high pressure chamber of a diamond anvil cell. The pressure dependence is found to be $d D/ dP = 1.458(6)$ MHz/kbar. However, since bulk diamond was used, the pressure measured was the averaged value over the sample volume. For studying effects such as the pressure-driven solidification process, a much higher spatial resolution is needed. Moreover, due to the presence of the bulk sensor, disturbance to the pressure distribution inside the medium may not be negligible. Therefore, in this work, we used a large quantity of nanodiamonds (ND), as opposed to the usual case of single ND, as spatially-resolved sensors inside the pressure medium, and the particle sizes are negligible compared to the pressurized region, which should minimize the disturbance to the actual pressure distribution. One may also want to use implanted NV\textsuperscript{-} centers to perform a similar study, but it is crucial to keep in mind that the pressure felt by the implanted NV\textsuperscript{-} centers is dominated by the uniaxial pressure at the tip of the anvil. Besides, one should be very careful about the relative shift between the high-pressure chamber and the implanted NV\textsuperscript{-} centers at the tip of the anvil. Furthermore, NDs are compatible with any kind of transparent anvil, but implanted NV\textsuperscript{-} centers work only for high quality diamond anvil.

Fig.~\ref{fig1}(c) shows the schematic drawing of our pressure cell configuration utilizing a microcoil setup. \cite{Alireza03} This microcoil setup has been used for the high-pressure measurement of the magnetic susceptibility, \cite{Alireza03,Klintberg12,Yip17} nuclear magnetic resonance, \cite{Haase09} the de Haas-van Alphen effect \cite{Goh08} and, more recently, ODMR spectra. \cite{Yang18} The pressure medium is confined by a metallic gasket made of beryllium copper, while two aligned moissanite anvils are used for generating pressure. The central hole of the gasket is 400 \textmu m , and it contains a sample, a microcoil, NDs and a ruby inside. To optimize the microwave (MW) throughput as well as minimizing heating, we use a 200 \textmu m diameter microcoil to transmit MW for ESR measurements. In order to probe the local pressure, NDs were drop-casted either on the top of one anvil surface all over the pressurized region or on the surface of a dummy sample. Here we used 1 \textmu m NDs with nitrogen concentration 3~ppm. It contains hundred thousands NV\textsuperscript{-} centers inside one particle. There are four possible NV\textsuperscript{-} orientations in a diamond lattice, one of which is drawn in Fig.~1(a). All those NV\textsuperscript{-} centers in NDs are approximately evenly distributed among these four orientations. Due to the tiny aperture (pressure cell opening is around ~4 mm) restricted by the anvil cell design, the effective numerical aperture (\textit{N.A.}) is reduced to around ~0.1, which leads to an optical spatial resolution in the order of microns. By using NDs smaller than this, we can reach better spatial resolutions.

Two optical setups were used in this work to measure the ODMR spectrum. Both setups use 520 nm laser diodes to initialize and readout NV\textsuperscript{-} center's electron spin states. The setup shown in Fig.~1(d) is a simple fluorescence microscopy setup. The red fluorescence due to the phonon side band emitted by NV\textsuperscript{-} centers inside the laser illuminated spot were collected by a photodiode. It has poorer spatial resolution compared with the confocal setup. However, its performance is comparable to the typical ruby spectroscopy setup used in high pressure community. Here, we use this setup to benchmark our new method with the traditional ruby method. On the other hand, the confocal setup shown in Fig.~1(e) has good fluorescence collection efficiency and spatial resolution with which ODMR of NV\textsuperscript{-} centers in individual NDs can be measured with high contrast. The confocal setup has rarely been used in studying pressurized media before. Here we show the advantages of using this setup in various scenarios.

In this work, we chose Daphne oil 7373 as the pressure medium. As Daphne oil 7373 is one of the most widely used and studied media, we are able to compare our new methodology with the previously reported results. Besides, Daphne oil 7373 is known to solidify at around 20 kbar. \cite{Haga09,Aoyama07,Aoyama08,Matsui10,Kakurai08,Canfield15} This modest critical pressure provides a convenient platform for us to demonstrate the power of our method to study the solidification process.

\begin{figure}[t]
    \centering
        \includegraphics[width=0.45\textwidth]{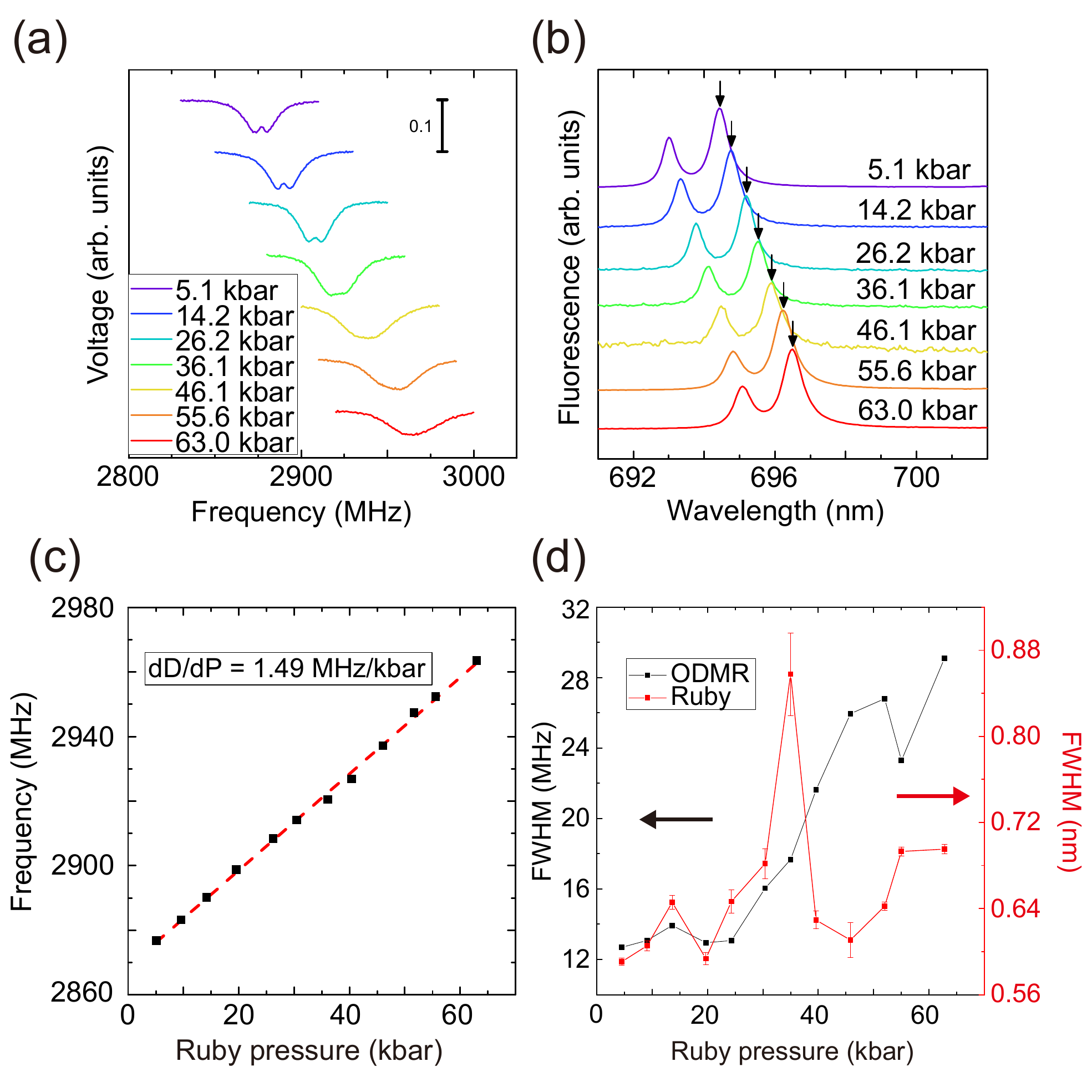}
    \caption{Benchmarking ODMR method with traditional ruby method. (a, b) shows the selected ODMR spectrum traces (ruby spectra) under different pressure. For the ODMR spectra, the center frequency of two resonances shifts to higher frequencies with higher pressures, while for the ruby spectra, the evolution of the \textit{R}$_1$ peak is used to determine the pressure. The pressure reported in legend are determined by ruby spectra. (c) The longitudinal ZFS \textit{D} against the pressure. $dD/dP$ = 1.49 MHz/kbar is perfectly linear over the whole pressure range even after solidification. (d) The full width at half maximum (FWHM) of ODMR (black) and ruby spectra (red) against the pressure. Both data show an increasing trend with the pressure.}
\label{fig2}
\end{figure}

\section{Benchmarking the ODMR spectroscopy with the ruby spectroscopy}
The structure and the simplified energy levels of NV\textsuperscript{-} centers in diamond are shown in Fig.~\ref{fig1}(a) and \ref{fig1}(b) respectively. There are four possible NV\textsuperscript{-} orientations depending on the relative position of the nitrogen and vacancy. We define the NV\textsuperscript{-} axis as the z-axis and the relevant ground state zero-magnetic-field Hamiltonian is given by
\begin{eqnarray}
H_{ZF} = D S_{z}^{2} + E \left( S_{x}^{2} - S_{y}^{2} \right),
\label{eq:ZFH}
\end{eqnarray}
where the longitudinal ZFS \textit{D} = 2.87 GHz at ambient pressure and room temperature. Depending on the nature and quality of the diamond sample, the transverse ZFS \textit{E} can range from negligibly small to several tens of MHz. In principle, the change of pressure can be derived from the change of \textit{D}. Similar to what happens in other pressure sensors, the change of \textit{D} can come from the change of pressure and temperature, as both of them affect the lattice parameters. In other systems, independent temperature sensors have to be used to decouple the effect of temperature. For NV\textsuperscript{-} centers, it is possible to distinguish temperature and pressure from its spectrum. $D$ under certain temperature can be calibrated when there is no pressure applied. In fact, at room temperature, $dD/dT=-74$ kHz/K and changes only slightly across a large temperature range. \cite{Guo11, Budker10, Hollenberg14} Below 20~K,  $dD/dT$ is nearly 0. Thus, ODMR spectra offers an avenue to determine the pressure with a much higher accuracy. In this section, we measure $dD/dP$ in Daphne oil 7373 and across the solidification region.

To determine the value of $dD/dP$,  we spread dense 1\textmu m NDs onto one of the anvil surfaces and used the fluorescence microscopy setup in Fig.~1(d) to measure the spectra at room temperature. The ruby spectra were also measured in the similar condition to calibrate the pressure. Fig.~\ref{fig2}(a) (Fig.~\ref{fig2}(b)) shows the selected ODMR traces (ruby spectra) under different pressures. For the ODMR spectrum, the center frequency of two resonances shifts to a higher frequency with a higher pressure, while for the ruby spectra, the evolution of the \textit{R}$_1$ peak is used to determine the pressure. Fig.~\ref{fig2}(c) shows the longitudinal ZFS \textit{D}, extracted from the center frequency of the resonances from ODMR, against the pressure. $d D/d P = 1.49 \pm 0.02$ MHz/kbar fits perfectly over the whole pressure range.

We summarized the previously reported $d D/d P$ in Table~\ref{table:NVcenter}. There are discrepancies among the results from different experiments and theories. Our results are in good agreement with two of the experimental works. \cite{Prawer14,Yang18} The slight difference in values may come from the variation in pressure calibration, such as the conditions of ruby used and the potential fluctuation of the room temperature. In these three works, four different pressure media have been used. As expected, the value of $d D/d P$ is pressure medium independent. Both in this work and in Doherty, \textit{et al.}\cite{Prawer14} the value of $dD/dP$ was measured across the solidification region. No measurable change has been observed. That means we can use $dD/dP\sim $1.49~MHz/kbar to calibrate the pressure across the solidification transition.

\begin{table}[h]
\centering
 \begin{tabular}{c c c c}
 \hline \hline
   & $d D/d p$ (MHz/kbar) & Pressure medium\\
 \hline
 Steele \textit{et al.}\cite{Curro17} & 1.172 $\pm$ 0.068 & Daphne 7373\\
 Doherty \textit{et al.}\cite{Prawer14} & 1.458 $\pm$ 0.006 & NaCl / Ne \\
 Yip \textit{et al.}\cite{Yang18} & 1.45 & Glycerol \\
 Ivady \textit{et al.}\cite{Gali14} & 1.030 & Theory\\
 This work & 1.49 $\pm$ 0.02 & Daphne 7373\\
 \hline \hline
 \end{tabular}
 \caption{The summary of the $d D/d P$ for pressure sensing using NV\textsuperscript{-} center. The data reported in this work are closed to the experimental data reported in \cite{Prawer14,Yang18}.}
 \label{table:NVcenter}
\end{table}

In Fig.~2(a), a strong increase of linewidth versus pressure can be seen. Since we kept the same MW power, this increase is not due to the power broadening. Furthermore, this broadening has a strong pressure dependence. We plotted the linewidths of NV\textsuperscript{-} center ESR spectra and ruby optical spectra versus the applied pressure in Fig.~2(d). The increase of linewidth starts around 26 kbar. Here both spectra are collected from sizable areas: the size of ruby is around 40~\textmu m while the laser spot size in the ODMR setup is about 6 \textmu m in diameter. The spatially inhomogeneous pressure distribution leads to the increase of linewidth. The critical transition pressure of 26 kbar agrees well with previous report of Daphne oil 7373's solidification transition at around 20 kbar. \cite{Haga09,Aoyama07,Aoyama08,Matsui10,Kakurai08,Canfield15} We note again that the slope in Fig.~2(c) does not depend on the solidification transition. We performed two linear fits for the region below and above 22 kbar, \cite{Aoyama07} and both values are nearly the same. As discussed above, these similar values show that the average pressure in the pressure medium does not get affected by the solidification process.

\section{The pressure induced solidification at room temperature}

\begin{figure}[t!]
    \centering
        \includegraphics[width=0.5\textwidth]{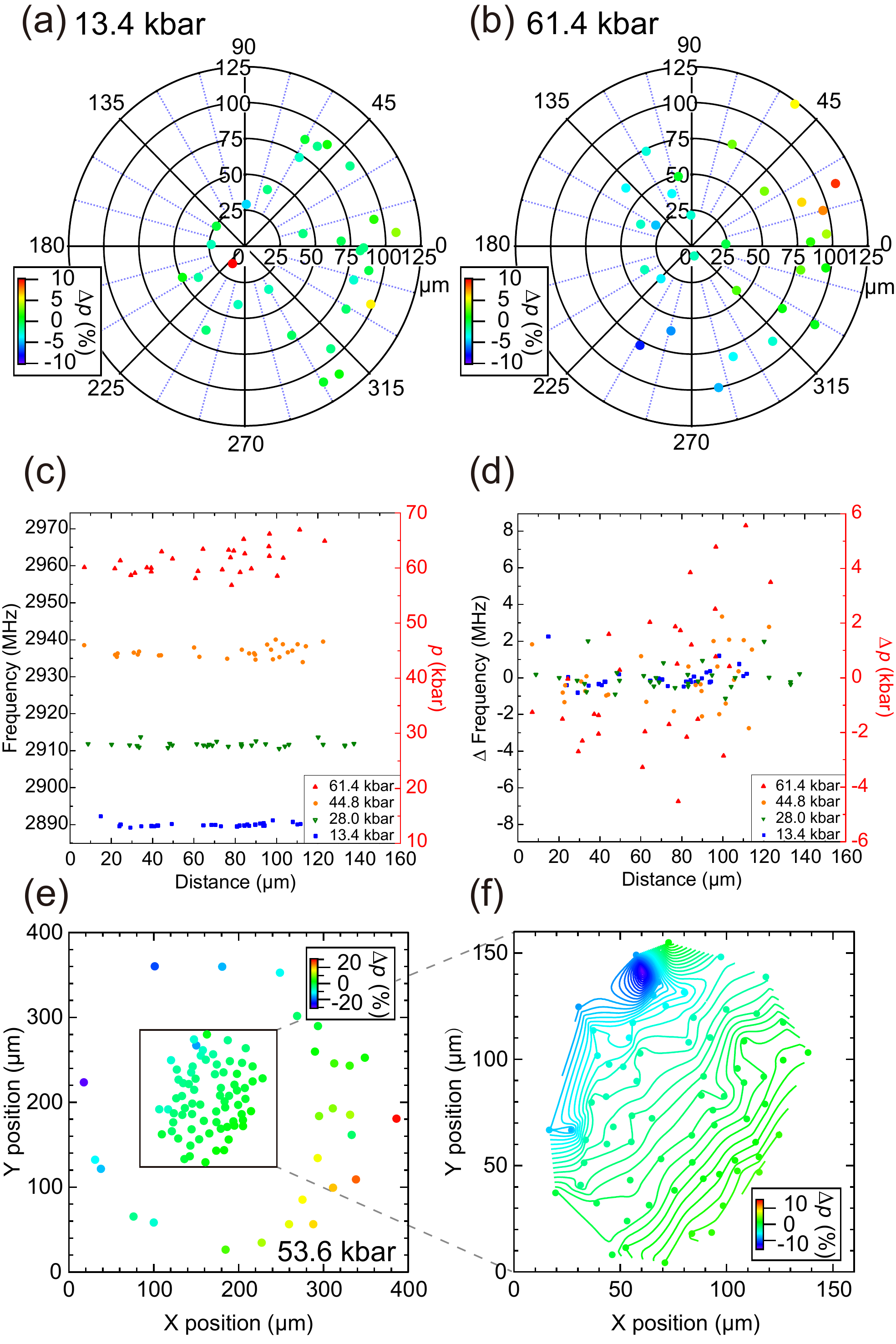}
    \caption{The pressure distribution mapped via NV\textsuperscript{-} centers. (a, b) show the polar plot of individual NDs with the color scale representing local pressure under average pressure 13.4 kbar and 61.4 kbar. The origin is defined as the center of the gasket hole. In the range of $\pm$10\%, no noticeable pressure gradient is observed under 13.4 kbar, while an obvious pressure gradient is observed under 61.4 kbar. Nonetheless, no circular pressure dependence is observed. (c, d) show the ZFS \textit{D} and the difference of \textit{D} against the distance from the origin. The data diverge at high pressures as a result of the linear gradient. It is clear that 61.4 kbar has a large pressure gradient. (e, f) is another dataset from a similar measurement in a pair of anvils with a slight angular offset between two anvil culets. (e) shows the pressure distribution of individual NDs with color scale representing local pressure at average pressure 53.6 kbar inside the gasket hole. Significantly huge pressure gradient is observed in the range of $\pm 30\%$. (f) is a zoom-in plot of (e), showing the contour plot inside the microcoil. A noticeable pressure gradient is observed in the range of $\pm 15\%$.}
\label{fig3}
\end{figure}

Measuring the local pressure with a confocal microscopy setup provides the resolution of around 1 \textmu m, the best spatial resolutions so far, for studying inhomogeneity in the process of solidification under pressure. It can also measure the position of the NDs precisely. Spatial maps of nearly 30 NDs from confocal scans are shown in Fig.~\ref{fig3}(a, b). The center of the plot is the center of the gasket hole. The spots are the NDs, which are randomly distributed in the pressurized region. The pressure sensed by each ND is determined by ODMR spectra with the slope chosen to be 1.49~MHz/kbar as determined in the previous section. The average pressure is 13.4 kbar (Fig.~3(a)), 61.4 kbar (Fig.~3(b)). The local pressure can be determined with high precision. The color in the figure shows the relative offset from the average value. As expected, the pressure inhomogeneity is much higher in 61.4 kbar than 13.4 kbar, as it is in the non-hydrostatic condition.

A pressure inhomogeneity study via the photoluminescence of a bulk GaAs sample with a spatial resolution of several hundreds \textmu m was reported before. \cite{Neu88} Here we present a more detailed 2-D map of a pressure medium using 1 \textmu m NDs. The spatial pressure distribution in non-hydrostatic condition can give hints on how solidification starts. Because the gasket and anvils are circularly symmetric, it was speculated that the pressure distribution would follow the same symmetry. In many works, the pressure distributions were plotted in radial directions. \cite{Block73,Aoyama08} However, the pressure distribution in Fig.~3(b) is closer to a linear dependence than a circular one. We measured the pressure distributions in a series of pressures across the solidification transition. The local pressure dependence on its distance to the center of the gasket hole is plotted in Fig.~3(c). There is no obvious sign of the circular symmetry. Furthermore, we can see the pressure is more uniform in the center than near the edges. This overall linear pressure distribution can come from the way the pressure is changed: two screws on opposite sides of the cell body are tightened in small steps one after the other. This can break the circular symmetry and give the linear gradient. We check this hypothesis with another pressure cell with the similar design in Fig.~5, the direction from high pressure region to low pressure region is nearly parallel to the line joining the two screws, as we predicted. The linear dependence can also come from the other imperfections, for example there can be a small angle between the culets of two anvils, which can also explain why the pressure is more non-uniform near the edges. This can be seen in another measurement, shown in Fig.~3 (e, f). Although great care has been taken during pressure cell assemblies, especially the alignment of two anvils, a slight angular offset between two anvil culets remains, i.e. one anvil is slightly tilted relative to another anvil. As a result, a strong ($\pm30\%$) pressure inhomogeneity was observed. This shows the importance of anvils alignment. Again from this measurement, we can confirm the two observations from the measurements in Fig.~3: the overall linear spatial pressure distribution, more uniform in pressure in the cell center and wider pressure differences near the edges. 

\begin{figure}[!t]
    \centering
        \includegraphics[width=0.5\textwidth]{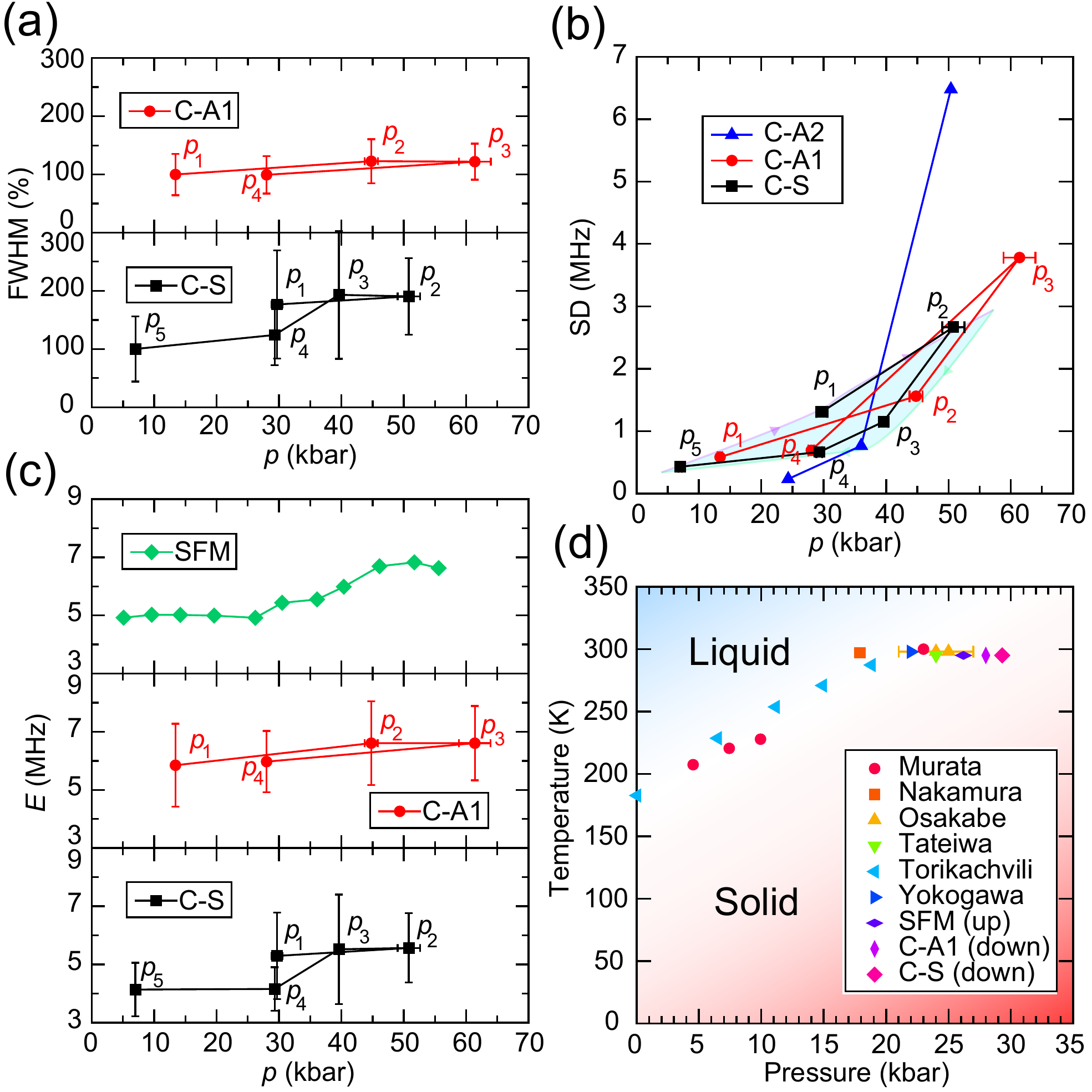}
    \caption{The pressure-driven solidification process. (a) shows the FWHM of spectra measured in C-A1 and C-S (definition of these abbreviations is explained in the main text). Data of C-A1 show a gradual increase with pressure, while those of C-S have a much larger increase. (b) shows the SD of C-A2, C-A1, and C-S. Both data show a significant increase at $\sim$ 30kbar. We can determine the onset of the critical pressure of Daphne oil 7373 being 28 kbar in C-A1 and 29.3 kbar in C-S, which is slightly higher than the previously reported value. (c) shows the ZFS \textit{E} of C-A1 and C-S. These behaviour are similar to (a). Data of SFM are plotted as an additional information, which shows a sharp turn at 26 kbar. (d) shows the temperature-pressure phase diagram of Daphne oil 7373. We also include the previously reported results and the results of this work.}
\label{fig4}
\end{figure}

The inhomogeneous pressure distribution in solidified media can be from a spatial stress gradient and a local shear stress. Since NV\textsuperscript{-} center is a point defect with angstrom size, using single NV\textsuperscript{-} center to probe the pressure distribution would give the best resolutions in both. Unfortunately, our current cell design is optimized for transport but not optical measurements, which limits our ability to do single NV\textsuperscript{-} center ODMR measurement. In this work, we used single ND with a few hundred thousands NV\textsuperscript{-} centers distributed randomly within 1 \textmu m. This averaging effect can lead to a line broadening effect. It has been known that the change of linewidth is a good indicator for the medium freezing effect. In general, inhomogeneity due to non-hydrostatic medium can lead to spectrum broadening in resonance type experiments. Indeed, various groups have already used linewidth changes as such an indicator. \cite{Haga09,Kakurai08,Jacobsen07,Aoyama08} In the fluorescence microscopy measurements in the previous section, the spatial resolution is beyond 6 \textmu m, the spectra broadening shown in Fig.~2 (d) are mainly due to the spatial inhomogeneous pressure distribution. This can be verified by examining the data in Fig.~3 (d), which is the collection of local pressure offsets relative to the average pressure in the cell. The pressure offset distribution in Fig.~3 (d) is in agreement with the linewidth in Fig.~2 (d).

The onset of the spectrum broadening can be the signature of the onset of the solidification process. Nevertheless, we introduce three physical parameters as indicators to determine the critical transition pressure. The first indicator is the linewith of single ND's ODMR spectrum. The broadening should come from the pressure spatial inhomogeneity within particle size. Here we use the relative linewidth, i.e. the linewidth relative to the linewidth at the lowest pressure, to eliminate the power broadening effect. Motivated by Klotz \textit{et al.}, \cite{Marchand09} we use the standard deviation (SD) of the pressure measured in various locations in the cell as the second indicator, which shows the pressure inhomogeneity in the whole pressurized region. The third indicator is the transverse ZFS \textit{E} of single ND particle, as it shows the strain inside the particle corresponding to the local inhomogeneity of the pressure distribution.

We measured these three quantities in four different measurements in Fig.~4(a-c): C-A2 ("C-A" stands for "confocal setup and NDs on the anvil") is the measurement shown in Fig.~3(e, f), SFM (stands for simple fluorescence microscopy) is the measurement shown in Fig.~2, C-A1 is the one in Fig.~3(a - d), C-S (stands for "confocal setup and NDs on the sample") is similar to C-A1 except the NDs were drop-casted on a dummy sample instead of on an anvil. 
Since each time when the pressure is modified there are permanent changes within the cell, it is important to track the sequence of the applied pressure, as marked in the figures as "\textit{p}\textsubscript{1}, \textit{p}\textsubscript{2},...". All three plots show quantitative agreements on the pressure dependency. It is interesting to see the similar broadening effect macroscopically (SFM) and microscopically (C-A1). It indicates that during the solidification process, the pressure inhomogeneity happens in both macroscopic scale as well as sub micron scale, and with strong angular dependence as shown in the change of strain in Fig.~4(c). Thus, our results show that NV\textsuperscript{-} centers can be employed as a highly sensitive tool to study pressure uniformity.

The onset critical pressure in these three methods, which is around 28 kbar (26.2, 28, and 29.3 kbar respectively), are in good agreement. By carefully examining the sequence of the applied pressure, it is interesting to notice that the pressure inhomogeneity (critical pressure) is higher (lower) during the process of increasing pressure from freshly prepared cell, while the opposite behaviour has been observed for the process of pressure release from the maximum pressure. This bifurcation-like phenomenon can be explained as following: during the deformation of pressure cell (especially the gasket) with an increasing pressure, local pressure "hotspot" can be formed which triggers the solidification process at a lower pressure, while after relaxing the pressure in an extensively deformed cell, the pressure distribution can be more uniform. From this speculation, the critical pressure measured during pressure releasing should be closer to the true critical pressure of the pressure medium.

The critical pressure of the solidification phase transition is a function of both the temperature and pressure. We combined previously reported results and the results of this work to construct the phase diagram in Fig.~4(d). For the data from this work, since we noticed a discernible pressure hysteresis, we add this information in the figure legend as "up/down". There are several reasons for the slightly difference in the room temperature critical pressure among our results and others. Since the pressure of the cell can only be changed with a finite step, there is an intrinsic limitation on the resolution. On the other hand, as shown in this work, the detailed cell structure, the quality of the cell preparation and even the sequence of applying pressure can change the pressure distribution and affect the measured critical pressure dramatically. 

\section{The temperature dependence of the solidification process}

\begin{figure}[h]
    \centering
        \includegraphics[width=0.5\textwidth]{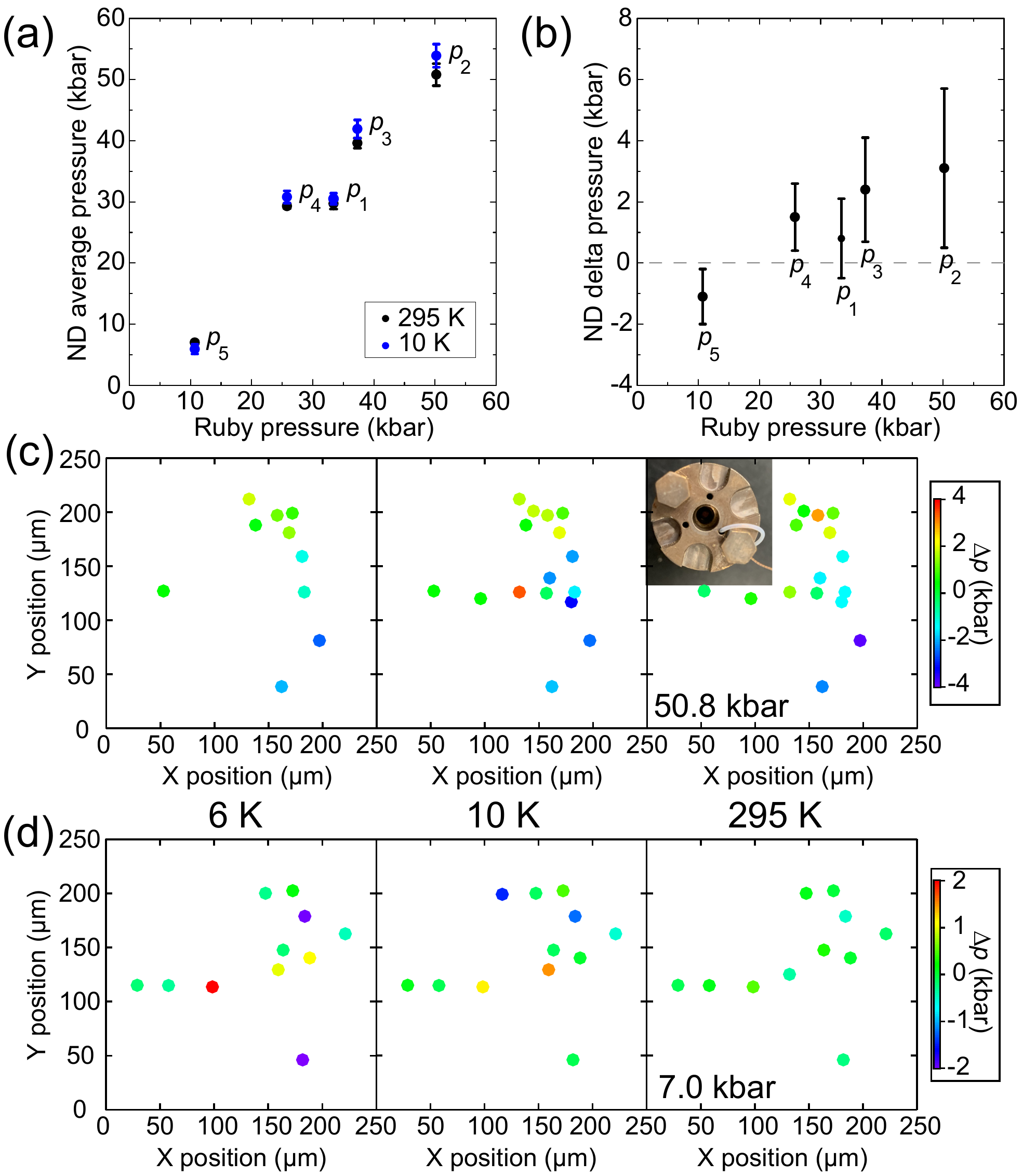}
    \caption{The change of pressure distribution with temperature. (a) shows the average pressures at room temperature and low temperature. The pressure in x-axis was measured by ruby spectra in room temperature. The pressure in cryogenic temperature is overall higher than the ones measured at room temperature. 
    (b) shows the pressure offsets in low temperature from room temperature. The pressure increases by nearly 5\% of the average pressure. (c, d) shows the spatial pressure distribution at different temperature of 50.8 kbar and 7.0 kbar. Pressure distribution changes are observed after cooling down, even when it is already in non-hydrostatic state in room temperature as shown in (c). A photo taken from the optical axis of the confocal setup is shown in the insert of (c). The direction from high pressure region to low pressure region is nearly parallel to the line joining the two screws for applying the pressure. On the other hand, when it is hydrostatic state at room temperature, inhomogeneity rises up at low temperature as shown in (d). No spatial pattern can be found in this scenario, it may due to the reason that the solidification starts randomly inside the medium.
    }
\label{fig5}
\end{figure}

A large portion of the high pressure research is taking place in cryogenic conditions, as that is the region where quantum effects matter. Thus, the temperature dependence is important to be measured. From the phase diagram in Fig.~4(d), it is clear that solidification happens for Daphne oil 7373 in cryogenic temperatures at any pressure. Here we perform the temperature dependency study by mounting the pressure cell in the cold-finger of a cryostat made by Montana Instruments. To study the pressure dependence at different temperatures, a decoupling of pressure and temperature effects is necessary. However, calibrating temperature in a confined device with possible temperature gradient is tricky. Usually an extra pressure insensitive temperature sensor is used for the temperature calibration. Fortunately for NV\textsuperscript{-} centers, there is nearly no temperature dependence of longitudinal ZFS \textit{D} below 30 K. Therefore, the change of \textit{D} is purely due to the change of pressure. We can thus compare the pressure distribution at 6-10 K with the results at room temperature.

We tracked more than 6 NDs within the pressurized region over different pressures and temperatures. We calculated both the average pressure over the whole region and the pressure distribution. The average pressures are shown in Fig.~5(a). The pressure in cryogenic temperature is overall higher than the ones measured at room temperature. In Fig.~5(b), we plot the offsets versus average pressure. The pressure increases by nearly 5\% of the average pressure. There were measurements showing pressure changes at low temperature. \cite{Aoyama07,Aoyama08,Shirakawa97,Canfield15} Specifically for Daphne oil 7373, it was reported that the pressure decreased at low temperature around 1.5 kbar in a piston cylinder clamp cell. \cite{Shirakawa97} The difference may come from the different designs of the pressure cells. By analysing the structure of our pressure cell carefully, we derived an increase of pressure in the order of a few kbar due to the different thermal contraction ratios of materials used for fabricating the cell, which fits the order-of-magnitude we measured. We point out that the actual offset can also come from other factors beside thermal contraction, such as the mechanical strength of the gasket. As can be seen in Fig.~5(b), the offset in \textit{p}\textsubscript{1} is smaller than \textit{p}\textsubscript{3} and \textit{p}\textsubscript{4}. This is because, after the plastic deformation under high pressure, the cell body has less resistance to the thermal contraction at the cryogenic temperature.

The spatial pressure distribution also changes after cooling down. This happens even when it is already in non-hydrostatic state in room temperature as shown in Fig.~5(c). Besides, when the cell is hydrostatic at room temperature, inhomogeneity rises up at low temperature due to freezing as shown in Fig.~5(d). No spatial pattern can be found in this scenario, it may due to the facts that the solidification starts pure randomly inside the medium.

\section{Pressure relaxation over time}

\begin{figure}[h]
    \centering
        \includegraphics[width=0.5\textwidth]{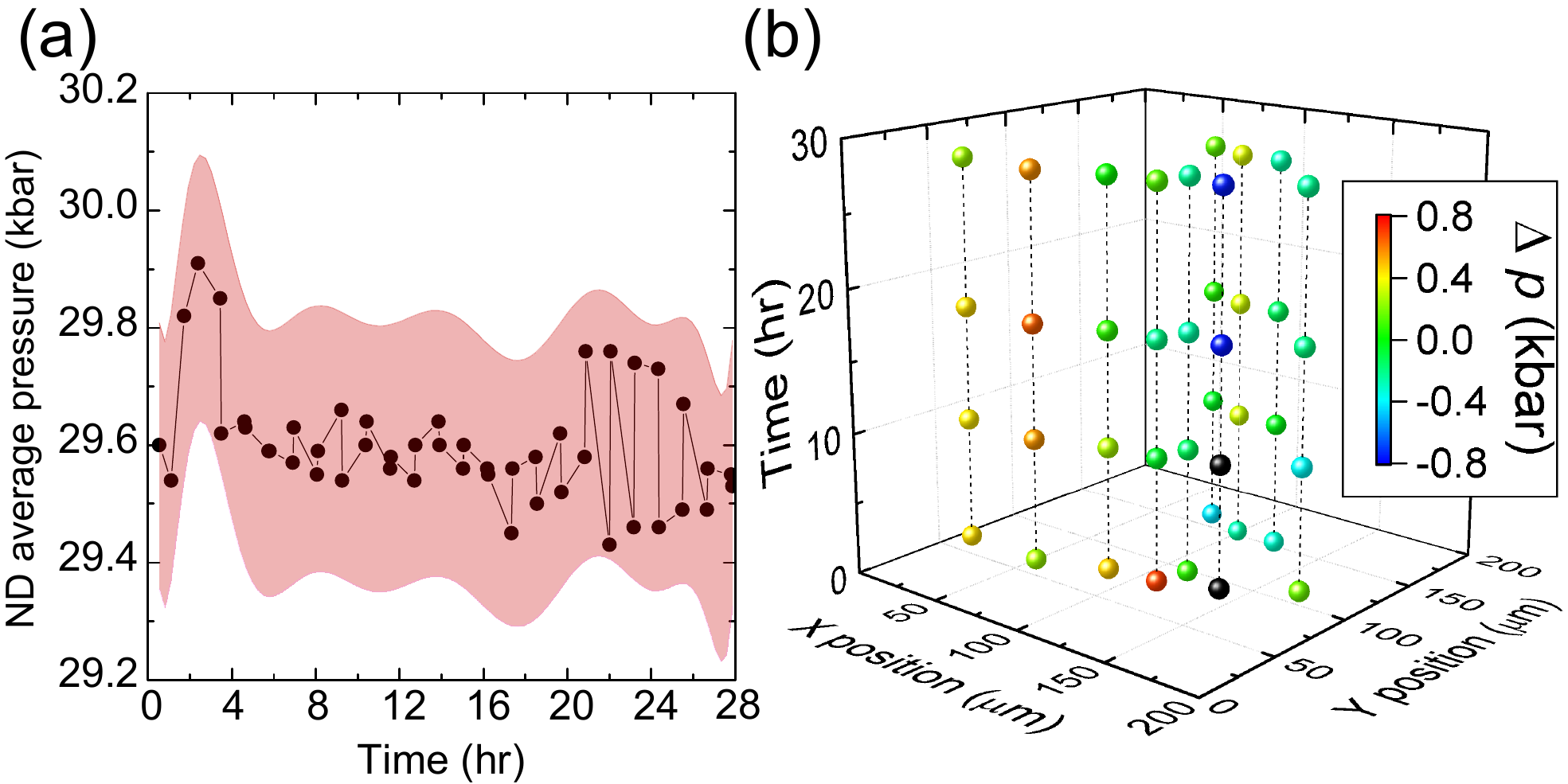}
    \caption{Pressure relaxation over time.  (a) shows the average pressure of 10 NDs over time. We define the starting time (t = 0) to be the moment we use a hydraulic press to change the cell pressure. The pink shadow shows the estimated error bar. (b) shows the pressure distribution of 10 NDs over time. The color of each point represents the pressure offset from the average pressure. The X-Y axes indicate the position of the NDs inside the pressurized region. The dashed lines connect the same NDs over time. The measurements of the NDs were performed one by one in the same order at different time.
    }
\label{fig6}
\end{figure}

The pressure of the cell is changed stepwise. It is important to know the process of pressure relaxation after the pressure suddenly changes. This process can again be monitored using NV\textsuperscript{-} centers in NDs. Here we tracked 10 NDs over a period of a day after we decrease the pressure from 39.6 kbar to 29.3 kbar at room temperature. For the measurement, we have taken extra care to calibrate the temperature to isolate the pressure effect from the temperature effect. Fig.~6(a) is the average pressure of 10 NDs against time. We define the starting time (t = 0) to be the moment when we use a hydraulic press to change the cell pressure. We can see that the pressure fluctuates and decreases quickly in 5-6 hours, and then slowly relax to the final pressure after nearly a day. Note that the pressure measured by ruby at the start of this pressure point and at the end also shows a pressure decrease of 1.2 kbar. On the other hand, as shown in Fig.~6 (b), the pressure inhomogeneity is also changed greatly in the time scale of 1 day.

\section{Conclusion}

We presented an experimental methodology and a comprehensive study of pressure distribution inside pressure cells under various experimental conditions with NV\textsuperscript{-} centers in NDs as the quantum sensor. The pressure dependency of ODMR spectrum is calibrated with the traditional ruby method and it is in good agreement with previous results. With this calibration, we mapped the pressurized region by multiple NV\textsuperscript{-} centers with high precision and spatial resolution, covering wide temperature and pressure ranges. With the capability to study the pressure environment, we studied both pressure driven and temperature driven solidification phase transitions. We found out that the average pressure and the pressure distribution are sensitive to experimental details, such as the mechanical details of the pressure cell, the sequence of applied pressure and the quality of the various cell components. Thus, for the measurements in which the pressure uniformity is crucial, for example quantum oscillations for Fermi surface mapping, the pressure inhomogeneity can be evaluated with high precisions. The method we presented here can give the unique power and compatibility to make the \textit{in-situ} pressure measurements together with other experimental probes. The protocol described in this work can also be used to study other pressure media. Moreover, since it has been demonstrated that NV\textsuperscript{-} center can also be used as an excellent magnetic field sensor in high pressure instrumentation, \cite{Yang18, Roch18, Yao19} together with temperature and electric field sensing ability, \cite{Lukin13, Wrachtrup13, Guo11, Budker10, Hollenberg14, Wrachtrup11, Yao18} the NV\textsuperscript{-} center in these tiny diamond particles can be used as a versatile, spatially resolved sensor for high pressure studies. Furthermore, a recent research on the 1042 nm zero phonon line of the NV\textsuperscript{-} center could also be adopted for a distinct approach to the low temperature high pressure measurements. \cite{Soltamov2017} Thus, NV\textsuperscript{-} center based sensors can be beneficial in different research areas especially in modern material research.

\begin{acknowledgments}
We thank Marcus W. Doherty and J. Wrachtrup for the fruitful discussions. We thank Kim Kafenda for technical supports. We thank S. Kasahara, Y. Mizukami, T. Shibauchi, and Y. Matsuda for providing samples. S.K.G. acknowledges financial support from Hong Kong RGC (GRF/14300418 and GRF/14301316). S.Y. acknowledges financial support from Hong Kong RGC (GRF/ECS/24304617, GRF/14304618 and GRF/14304419), CUHK start-up grant and the Direct Grants.
\end{acknowledgments}

\section*{Appendix: Parameter derivation from ODMR spectra of NV\textsuperscript{-} center}
Equation (\ref{eq:ZFH}) is a commonly used zero magnetic field Hamiltonian of NV\textsuperscript{-} center, which shows the dependence of pressure and temperature in the term \textit{D}, and strain and electric field in the term \textit{E}. With the presence of a magnetic field, an extra Zeeman splitting term will show up. From the ODMR spectrum, the ZFS \textit{D} is determined as follows:
\begin{eqnarray}
D = \frac{P_{R}+P_{L}}{2},
\end{eqnarray}
where $P_{R}$ is the right peak frequency and $P_{L}$ is the left peak frequency of the ODMR spectrum. Similarly, the ZFS \textit{E} is determined as follows:
\begin{eqnarray}
E = \frac{P_{R}-P_{L}}{2}.
\end{eqnarray}
The \textit{E} term is sensitive to the local strain so it is a useful indicator for the solidification process and pressure inhomogeneity, including spatial gradient and local shear stress.

Although this picture is widely accepted, one can have a much detailed formulation as discussed elsewhere. \cite{Yao18, Maletinsky19} Indeed, the Hamiltonian in Eq. (\ref{eq:ZFH}) is a simplified model. From Barfuss's work, \cite{Maletinsky19} the general spin-stress coupling involved three terms $H_{\sigma 0}$, $H_{\sigma 1}$ and $H_{\sigma 2}$. And the term $H_{\sigma 0}$ is also related to ZFS \textit{D}. \cite{Maletinsky19} We assumed in our case, which is under high pressure, that the change in \textit{D} is dominated by the change in pressure and the change in \textit{E} is dominated by the change in local shear stress and strain, so the Hamiltonian is reduced to Eq. (\ref{eq:ZFH}). The situation becomes more complicated if one also takes into account the electric field contribution. Depending on the quality of the diamond sample, the local charge environment can differ a lot. For instance, one possible factor of the large splitting of the doublet is a large random electric field induced by local environment of charges. \cite{Yao18}

In order to have a fair comparison of the relevant parameters upon pressure change, we have analyzed our ODMR spectra at all pressure points using two Lorentzians. This may be tricky because in some ODMR spectra, particularly at high pressure, the peaks merged into a single peak due to line broadening. However, even if a single broad peak is obtained, it has to be the result of merging two peaks. Using two Lorentz peaks fitting, we can recover the two individual peaks from the merged peak.

\end{document}